\documentclass[12pt]{iopart}
\usepackage{harvard}
\usepackage{graphicx}
\usepackage[normal]{subfigure}
\usepackage{booktabs}
\usepackage{lscape}
\usepackage{color}
\usepackage{epstopdf}
\usepackage{tikz}
\usepackage{gplpdf}
\usepackage{cellspace}

\newcommand{\otoprule}{\midrule[\heavyrulewidth]}
\begin{document}

\title[Charge exchange and ionisation in N$^{\mathbf{7+}}$, N$^{\mathbf{6+}}$, C$^{\mathbf{6+}}$ - H($n=1, 2$) collisions]{Charge exchange and ionisation in N$^{7+}$, N$^{6+}$, C$^{6+}$ - H($n=1, 2$) collisions studied systematically by theoretical approaches}

\author{K Igenbergs$^1$, J Schweinzer$^2$, A Veiter$^1$, L Perneczky$^1$, E Fr\"uhwirth$^1$, M Wallerberger$^1$, R E Olson$^3$ and F Aumayr$^1$}

\address{$^1$ Institute of Applied Physics, Vienna University of Technology, Association EURATOM-\"OAW, Wiedner Hauptstr.8-10/E134, A-1040 Vienna, Austria, EU\\
$^2$ Max-Planck-Institut f\"ur Plasmaphysik, Association EURATOM, Boltzmannstr. 2, D-85748 Garching, Germany, EU \\
$^3$ Physics Department, Missouri University of Science \& Technology, Rolla, MO 65401, USA}
\ead{igenbergs@iap.tuwien.ac.at}

\begin{abstract}

The introduction of gases like nitrogen or neon for cooling the  edge region of magnetically confined fusion plasmas has triggered a renewed interest in state selective cross sections necessary for plasma diagnostics by means of charge exchange recombination spectroscopy. To improve the quality of spectroscopic data analysis, charge exchange and ionisation cross sections for N$^{7+}$ + H($n=1,2$) have been calculated using two different theoretical approaches, namely the atomic-orbital close-coupling method and the classical trajectory Monte Carlo method. Total and state resolved charge exchange cross sections are analysed in detail.

In the second part, we compare two collision systems involving equally charged ions, C$^{6+}$ and N$^{6+}$ on atomic hydrogen. The analysis of the data lead to the conclusion that deviations between these two impurity ions are practically negligible. This finding is very helpful when calculating cross sections for collision systems with heavier not completely stripped impurity ions.

\end{abstract}

\submitto{\JPB}
\maketitle

\section{Introduction}

Charge exchange (CX) in collision processes between neutral hydrogen isotopes and  fully stripped ions has been the subject of a large number of studies in the past \cite{fritsch91,bransden80}. Renewed interest comes from thermonuclear fusion  research since cross sections for these processes are needed for a variety of applications, in particular charge exchange recombination spectroscopy (CXRS)~\cite{isler94}. 

CXRS is a standard plasma diagnostic tool to measure radial profiles of the ion temperature and density by using a heating or diagnostic neutral beam as the local source of neutral hydrogen isotopes inside the plasma. In principle CX from neutral hydrogen into any impurity ion species in the plasma can be used. However, best CXRS results are achieved for those with the highest abundance and using fully stripped ions facilitates the analysis. For many years fully stripped carbon was the dominant impurity in many fusion devices. Due to the change in choice of wall material replacing carbon with tungsten and/or beryllium (ASDEX Upgrade, JET and ITER) the abundance of carbon decreased~\cite{gruber09}. In addition, the use of metallic walls and in particular the divertor tiles in present and future fusion devices makes the intentional puff of impurity (seeding) gases mandatory in order to keep heat loads to exposed elements of the metallic wall below technical limits. Impurities like N, Ne and Ar are particularly good candidates to convert a considerable fraction of the heat flux into radiation leading to a more uniform power distribution over the inner walls of the fusion machine. This technique of controlling the heat exhaust in fusion devices with ITER- and reactor-relevant first walls is currently a focus of the worldwide fusion relevant plasma research. For applying the CXRS method in such plasmas the most promising multiply charged ions depend not only on intrinsic impurities (wall material), but also on the radiator used to cool the plasma edge~\cite{kallenbach10}.

For a mid-sized tokamak like ASDEX Upgrade it turned out that nitrogen has almost optimal radiation characteristics, because it radiates predominately in the plasma edge. As a positive surprise it turned out that the introduction of N does not only protect plasma facing components, but also improves significantly the performance of discharges~\cite{schweinzer11}. In order to understand this effect better absolute N$^{7+}$ density profiles are required. For the latter the knowledge of accurate CX cross sections is a prerequisite.

To calculate both charge exchange (CX) and ionisation (ION) cross sections, we applied the well-known atomic-orbital close-coupling (AOCC) method. Additionally we used the also well-known classical trajectory Monte Carlo (CTMC) approach to calculate cross sections for N$^{7+}$ + H($n=2$) since for this system the only available reference data comes from a scaling algorithm by~\citeasnoun{foster08}.

The neutral heating beam at current fusion experiments consists mainly of ground state D, but a fraction of the beam enters the plasma in an exited (metastable) state or becomes excited in collisions with plasma particles (electrons) when penetrating the plasma edge. Although this fraction is small, its contribution to the intensity of the observed spectral lines is substantial because the involved CX cross sections are about an order of magnitude larger than those for ground state hydrogen and electron capture from the excited state leads to population of higher n-shells, much closer to the observed transition lines in the visible range.~\cite{hoekstra98}

At impact energies above $\approx 30$ keV/amu, it is utterly important to take also ION into account, because it is strongly competing with CX. This is particularly the case for collisions with H($n=2$). Especially in our AOCC calculations it became necessary to include a large number of ionisation states on the hydrogen centre.

We will also briefly discuss $N^{6+}$ and C$^{6+}$ impacting on neutral ground and first excited state atomic hydrogen. Firstly, these two ion species are of interest in nuclear fusion research as well, secondly the comparison between the equally charged ions gives some insight into how calculations of highly charged, but not fully-stripped ions should be approached in the future.

We give a very short description of the used theoretical approaches and then focus on a detailed analysis of the calculated cross sections. Atomic units are used unless otherwise stated.

%%%%%%%%%%%%%%%%%%%%%%%%%%%%%%%%%%%%%%%%%%%%%%%%%%%%%%%
\section{Applied theoretical approaches}

We applied both the atomic-orbital close-coupling (AOCC) and the classical trajectory Monte Carlo (CTMC) treatments of the above stated collisional systems. Both are very well known and widely used to calculate inelastic cross sections. Thus, we will refrain from giving extensive explanations of the algorithms and only point out important details.
 
We applied the AOCC formalism~\cite{fritsch91,bransden92,gieler93} in its impact parameter description. In this semiclassical approach, the total electron wavefunctions are expanded into target- and projectile-centred traveling atomic orbitals (AO). The time-dependent Schr\"odinger equation is solved in the truncated Hilbert space following the procedure by~\cite{nielsen90,wallerberger11}. The appropriate choice of atomic basis states reflects the physical problem to be treated. The calculations are done in the so-called collision frame (impact parameter, $b$, parallel to x-axis, impact velocity, $v$, parallel to the quantization axis z). Real spherical harmonics are used for the angular part of the basis states. With this choice only states with the same symmetry with respect to a reflection on the scattering plane couple in the AOCC approach. Thus an initial state of positive reflection symmetry ($\ell = 0, m_{\ell}=0$) does not couple with states of negative reflection symmetry ($\ell = 1, m_{\ell} < 0$) and vice versa leading to basis sets of minimal size. In particular, the basis sets using the collision frame for calculations with initial states of negative reflection symmetry become small because all states with positive reflection symmetry can be omitted. The modus operandi is described in much greater detail in~\cite{igenbergs09}. For convergent calculations, we used basis sets consisting of up to 286 states on the ion centre ( $1 \leq n \leq 11$). Depending on the incident ion 54 (C$^{6+}$, N$^{6+}$) or 73 (N$^{7+}$) states on the hydrogen center were used. The 54-state basis consists of 20 pure hydrogen states ($1 \leq n \leq 4$) and 34 unbound pseudostates representing the ionisation continuum. The 73-state  hydrogen basis is made of 10 pure hydrogen states with Laguerre-type radial parts ($1 \leq n \leq 3$) and 63 unbound pseudostates. The used bases differ because for N$^{7+}$ + H more pseudostates were necessary to achieve converging results for ionisation cross sections.

For collisions of N$^{7+}$ with ground state H(1s), a number of experimental and theoretical studies have been performed and published by several groups \cite{meyer85,dijkkamp85,fritsch84,illescas99}. Additionally the Atomic Data \& Analysis Structure (ADAS) software package features a scaling algorithm (called ADAS315) that can quickly produce state-resolved CX cross section for any desired incident ion~\cite{foster08}. This algorithm also works for excited state H($n=2$) targets, but remained the only source we could compare our AOCC cross sections to since no other data for this collision system could be found in the literature. To resolve this unfavorable situation, we used the CTMC method as first outlined in~\cite{olson77} for exited state H targets to complement our AOCC calculations. 

Briefly, Hamilton's equations were solved for a mutually interacting three-body problem. A classical number $n_c$ is obtained from the binding energy $E_p$ of the captured electron relative to the projectile by

\begin{equation}
E_p = -\frac{Z^2_p}{2 n_c^2},
\end{equation}
where $Z_p$ is the charge of the projectile. Then, $n_c$ is related to the quantum number $n$ of the final state by

\begin{equation}
 \left [ (n-1) \left(n - \frac{1}{2} n \right) \right ] ^{1/3} \leq n_c \leq \left[ n (n+1) \left(n + \frac{1}{2} \right) \right] ^{1/3}.
\end{equation}

\noindent The cross section for a definite $n$-state is given by

\begin{equation}
\sigma _n = \frac{N(n) \pi b^2_{max}}{N_{tot}},
\end{equation}
where N(n) is the number of events of electron capture to the $n$-level and $N_{tot}$ is the total number of trajectories integrated. The impact parameter $b_{max}$ is the value beyond which the probability of electron capture is negligibly small.

%%%%%%%%%%%%%%%%%%%%%%%%%%%%%%%%%%%%%%%%%%%%%%%%%%%%%%%
\section{Results for N$^{\mathbf{7+}}$ + H($\mathbf{n=1,2}$)}

\begin{figure}[htbp]
\centering
\subfigure[\label{f:tot-a}]{\includegraphics[width=.48\columnwidth]{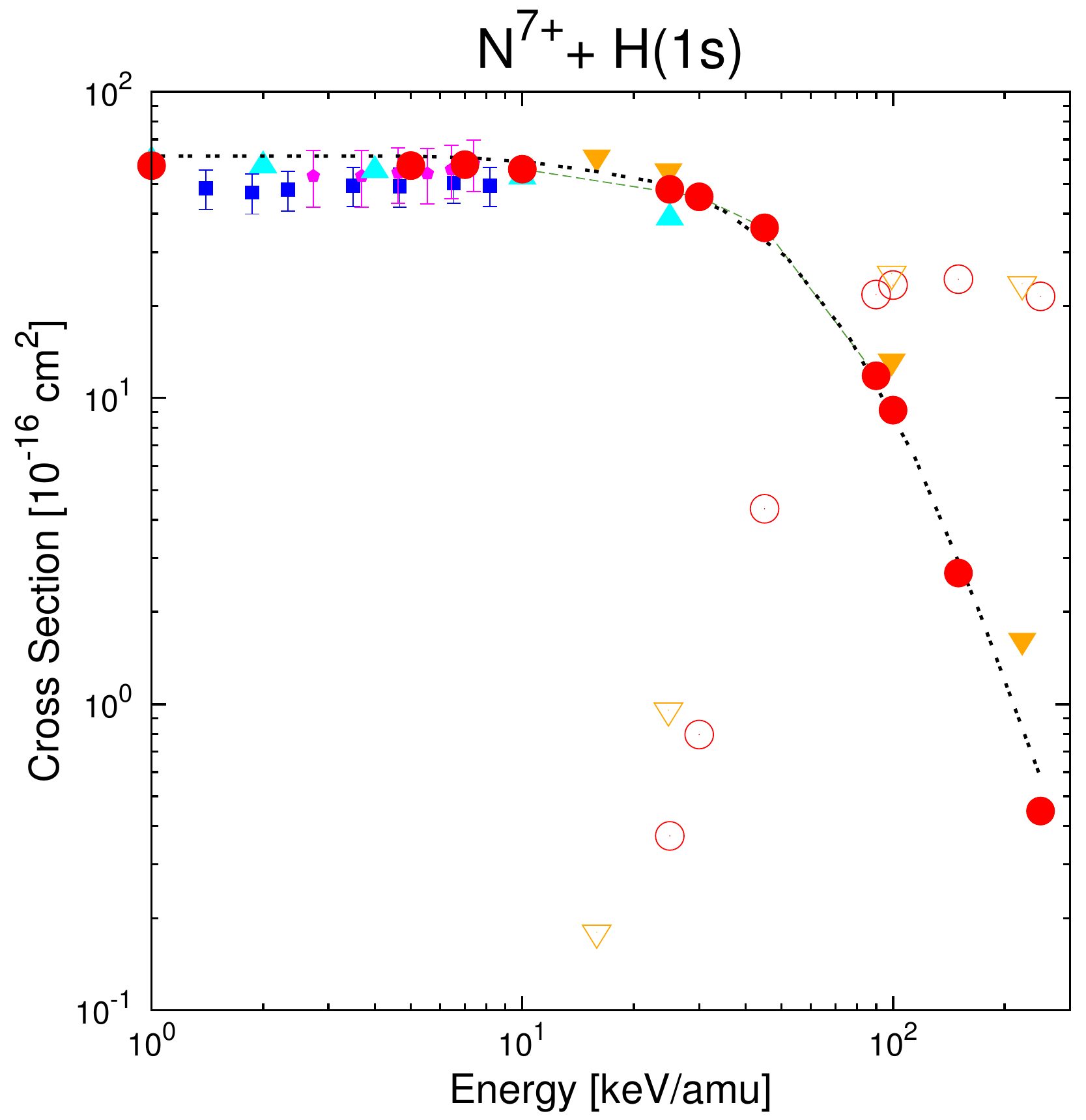}}
\subfigure[\label{f:tot-b}]{\includegraphics[width=.48\columnwidth]{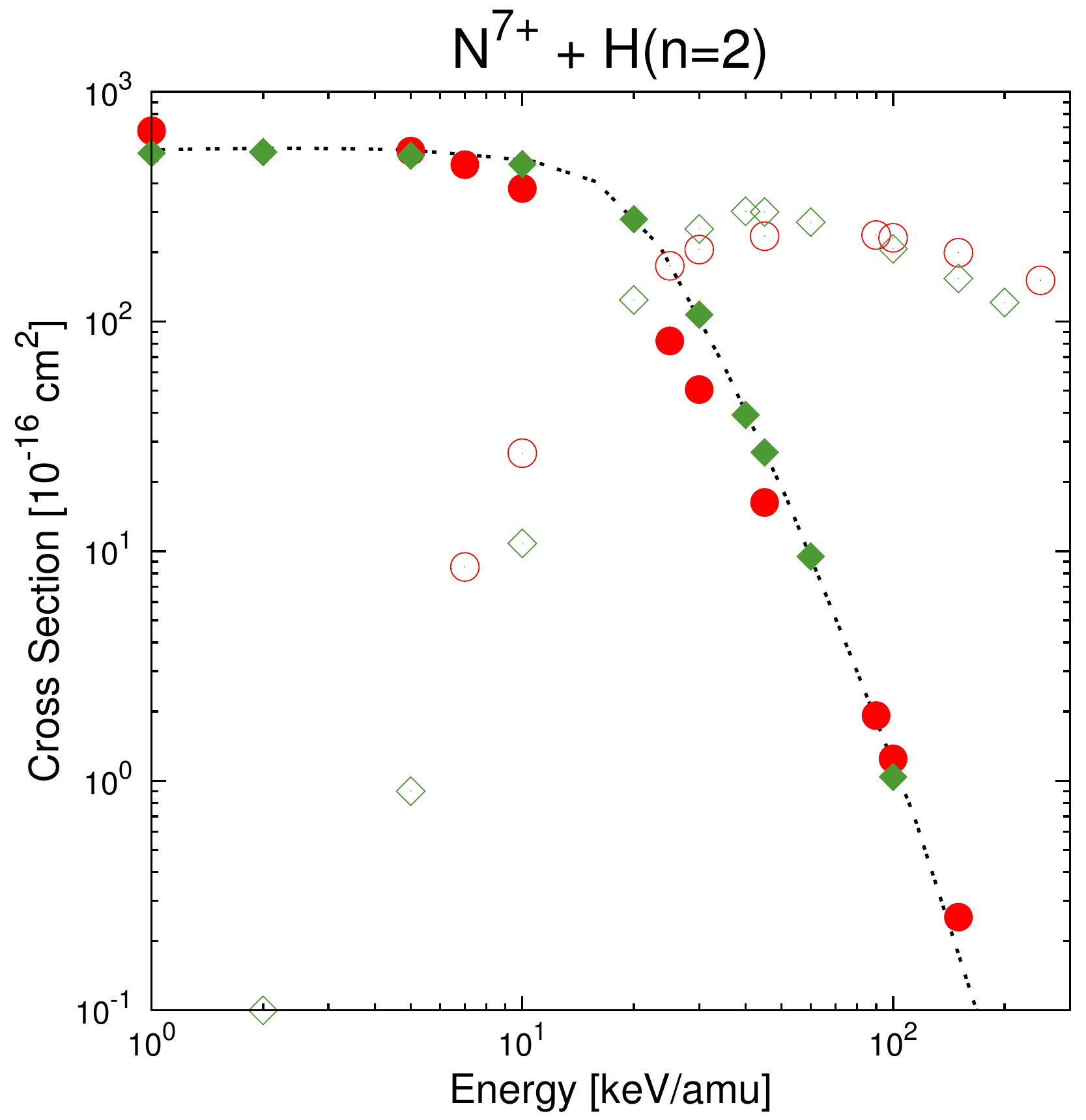}}
\caption{(Color online) Total cross sections for CX (full symbols) and ION (open symbols). (a) H(1s) target. Data presented in this work (AOCC \protect\gplps{7}\protect\gplps{6}, CTMC \protect\gplps{13}\protect\gplps{12}). For reference purposes we show experimental CX data from \protect\citeasnoun{meyer85} (\protect\gplps{5}) and \protect\citeasnoun{dijkkamp85} (\protect\gplps{15}) as well as results from various theoretical approaches: AO+ CX cross sections from \protect\citeasnoun{fritsch84} (\protect\gplps{9}), CTMC results for CX \mbox{(\protect\gplps{11})} from \protect\citeasnoun{illescas99}, scaled CX data (\protect\gplls{7}) calculated with ADAS315~\protect\cite{foster08} and ION (\protect\gplps{10}) again from \protect\citeasnoun{illescas99}. (b) H($n=2$) target.}
\label{f:tot}
\end{figure}

Fig.\ref{f:tot-a} shows total CX and ION cross sections for an H(1s) target, fig.\ref{f:tot-b} for an H($n=2$) target. The cross section in the latter case is the statistically weighted sum of the cross sections of the four different $n=2$ substates~\cite{igenbergs09}. 

We compare our results for the H(1s) target to experimental results from \citeasnoun{meyer85} and \citeasnoun{dijkkamp85} as well as AO+ CX cross sections from~\citeasnoun{fritsch84} and CTMC results for both CX and ION from~\citeasnoun{illescas99}. Additionally we show the previously mentioned scaled data from ADAS315 for both ground and excited state targets. For H(1s), the total CX cross sections (full symbols) agree well. The slight disagreement between our AOCC (\gplps{7}) and the AO+ calculations by \citeasnoun{fritsch84} (\gplps{9}) at E=25 keV/amu very likely originates from computational limitations of the basis size in 1984. For ionisation (open symbols)from H(1s) the agreement between AOCC cross sections and CTMC calculations from \citeasnoun{illescas99} is generally good, especially at higher energies, where ION plays an important antagonistic role to CX, the agreement is excellent.

\begin{figure}[htbp]
\centering
\subfigure[\label{f:resolvedn2-a}]{\includegraphics[width=.48\columnwidth]{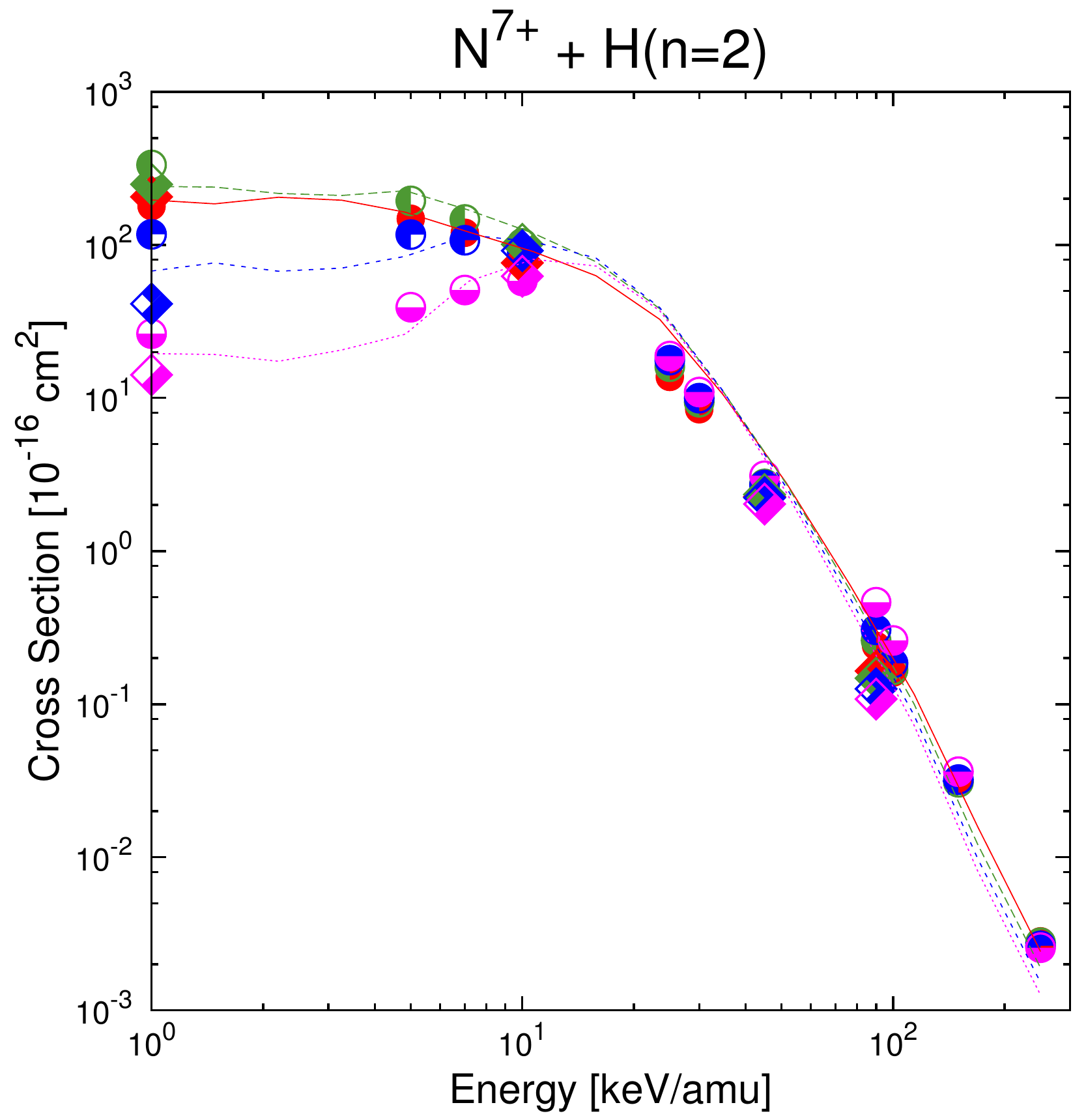}}
\subfigure[\label{f:nresolvedn2-b}]{\includegraphics[width=.48\columnwidth]{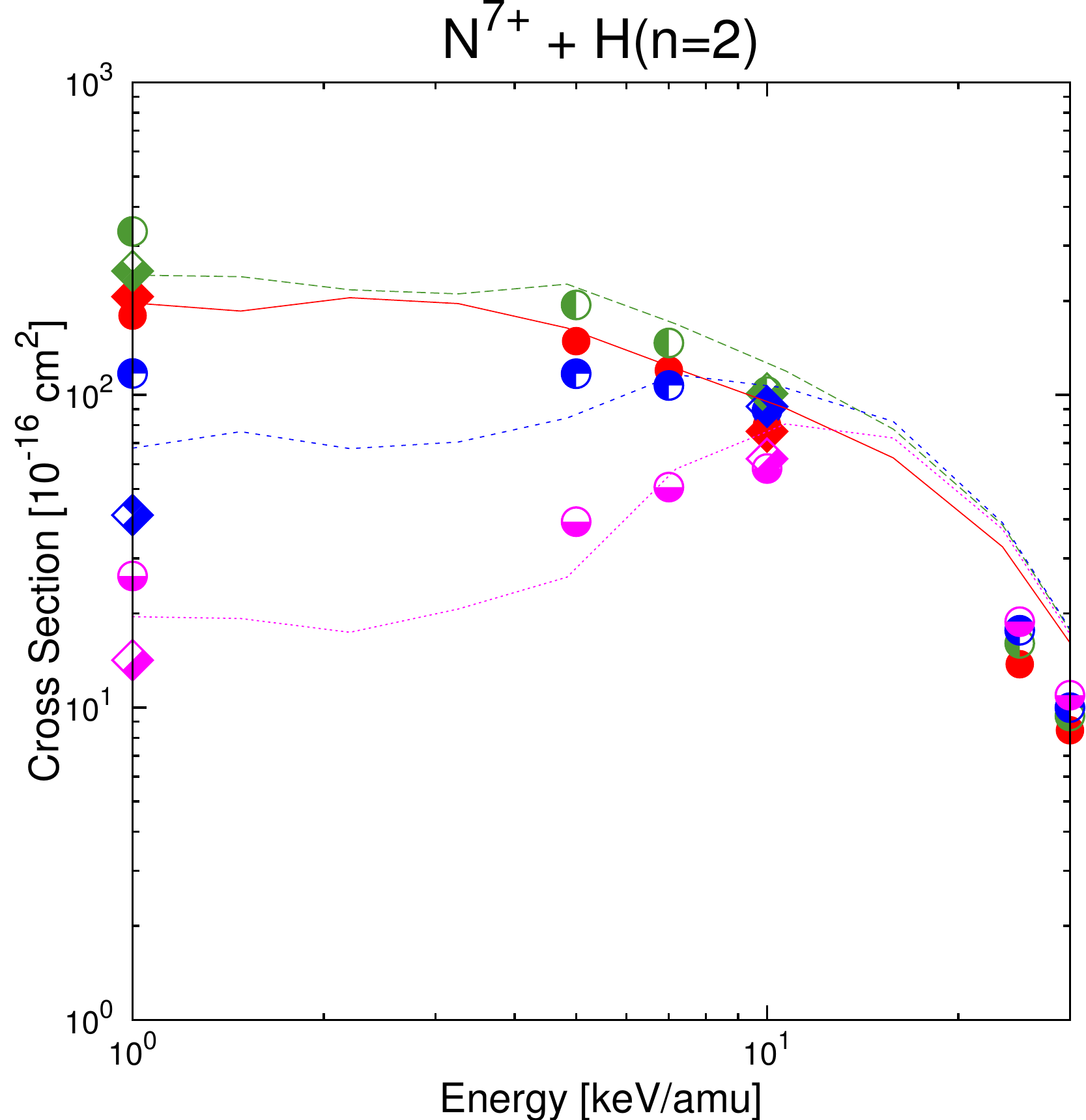}}
\caption{(Color online) $n$-resolved cross sections for electron capture from a H($n=2$) target. Circle symbols are used for AOCC data, rhombic symbols for CTMC, and lines for ADAS315~\protect\cite{foster08} data. $\sigma(n=8)$ (red,\protect\gplps{7},\protect\gplps{63},\protect\gplls{1}), $\sigma (n=9)$ (green,\protect\gplps{22},\protect\gplps{62},\protect\gplls{2}) , $\sigma(n=10)$ (blue,\protect\gplps{23},\protect\gplps{61},\protect\gplls{3}), and $\sigma (n=11)$ \mbox{(magenta,\protect\gplps{28},\protect\gplps{60},\protect\gplls{4})} (a) total energy range (b) 1-30 keV/amu. }
\label{f:nresolvedn2}
\end{figure}

The situation for the added up cross sections for an excited state H($n=2$) target is not as clear as for H(1s). AOCC, CTMC and ADAS315 data agree only reasonably well. But the AOCC data yield lower CX cross sections in the intermediate energy range between 10 and 45 keV/amu. ADAS315 uses CTMC calculations for other collision systems, e.g. C$^{6+}$ + H or O$^{8+}$ + H, interpolates between them and gives cross sections for - in this case - N$^{7+}$ + H($n=2$). So it comes with no surprise that these cross sections agree better with our CTMC calculations. Including higher lying $n$-shells in the AO basis does not improve the agreement of the total CX cross sections, because, as can be seen in fig.\ref{f:nresolvedn2}, the differences also occur in the $n$-resolved cross sections at $n < n_{max} = 11$. For ION from H($n=2$) targets the agreement between our AOCC and CTMC results is sufficiently good. The results start to diverge at high impact energies (E $\geq 150$ keV/amu), but in this energy range the CX cross sections are already several orders of magnitude lower that at the plateau between 1 and 10 keV/amu. Thus it is very inefficient to use neutral beams with these high energies for diagnostic purposes and the diverging trend at high energies will have a negligible influence on the evaluation of spectroscopic data.

\begin{figure}[htbp]
\centering
\includegraphics[width=.48\columnwidth]{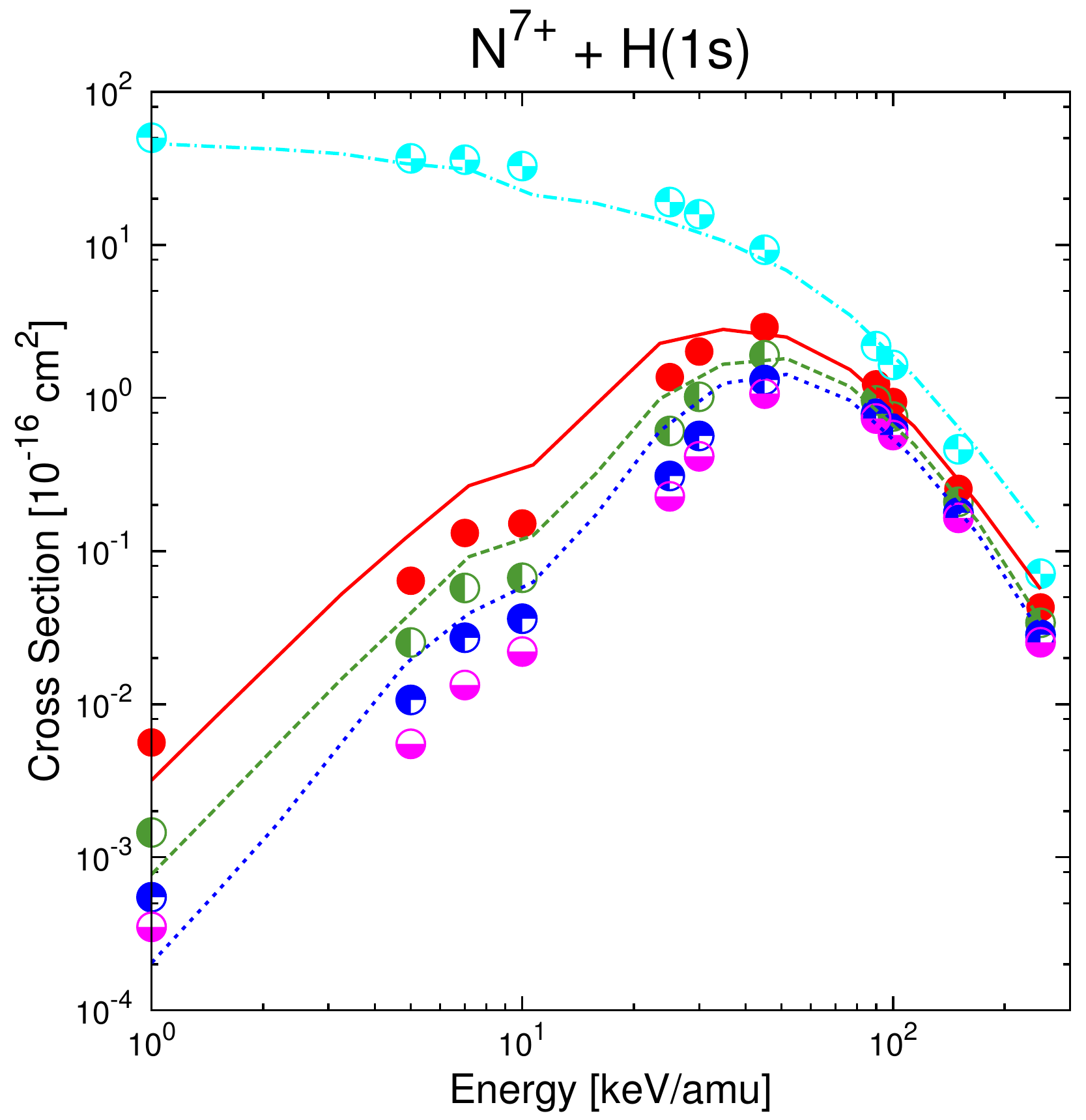}
\caption{Color online. $n$-resolved cross sections for electron capture from the H(1s) target. Circle symbols are used for AOCC data and lines for ADAS315~\protect\cite{foster08} data. $\sigma(n=5)$ (cyan,\protect\gplps{26},\protect\gplls{5}), $\sigma(n=8)$ (red,\protect\gplps{7},\protect\gplls{1}), $\sigma (n=9)$ (green,\protect\gplps{22},\protect\gplls{2}) , $\sigma(n=10)$ (blue,\protect\gplps{23},\protect\gplls{3}), and $\sigma (n=11)$ \mbox{(magenta,\protect\gplps{28}).}}
\label{f:nresolved1s}
\end{figure}

For the analysis of experimental CXRS data, state resolved cross sections are needed. \mbox{$n$-resolved} cross sections are shown in fig.\ref{f:nresolved1s} (H(1s) target) and fig.\ref{f:nresolvedn2} (H($n=2$) target). We only show $\sigma(n)$ for those shells that are either the main capture channel ($n=5$ for H(1s) target, $n=9$ for H($n=2$) target) or result in visible emission lines (i.e. N VII ($9\rightarrow8$) line) including contributing cascades. For the H($n=2$) target, the differences between our AOCC and CTMC calculations might be as large as a factor of 5 at low energies, but the agreement quickly ameliorates when approaching higher impact energies, i.e. those energies that are of importance for CXRS.  The results of our AOCC calculations are summarized in Table \ref{t:n7h1s} and \ref{t:n7hn2}, where total CX and ION as well as $n$- and $n\ell$-resolved CX cross sections are given for all calculated impact energies.

\begin{figure}
\centering
\subfigure[\label{f:n7hnl_1-a}]{\includegraphics[width=.49\columnwidth]{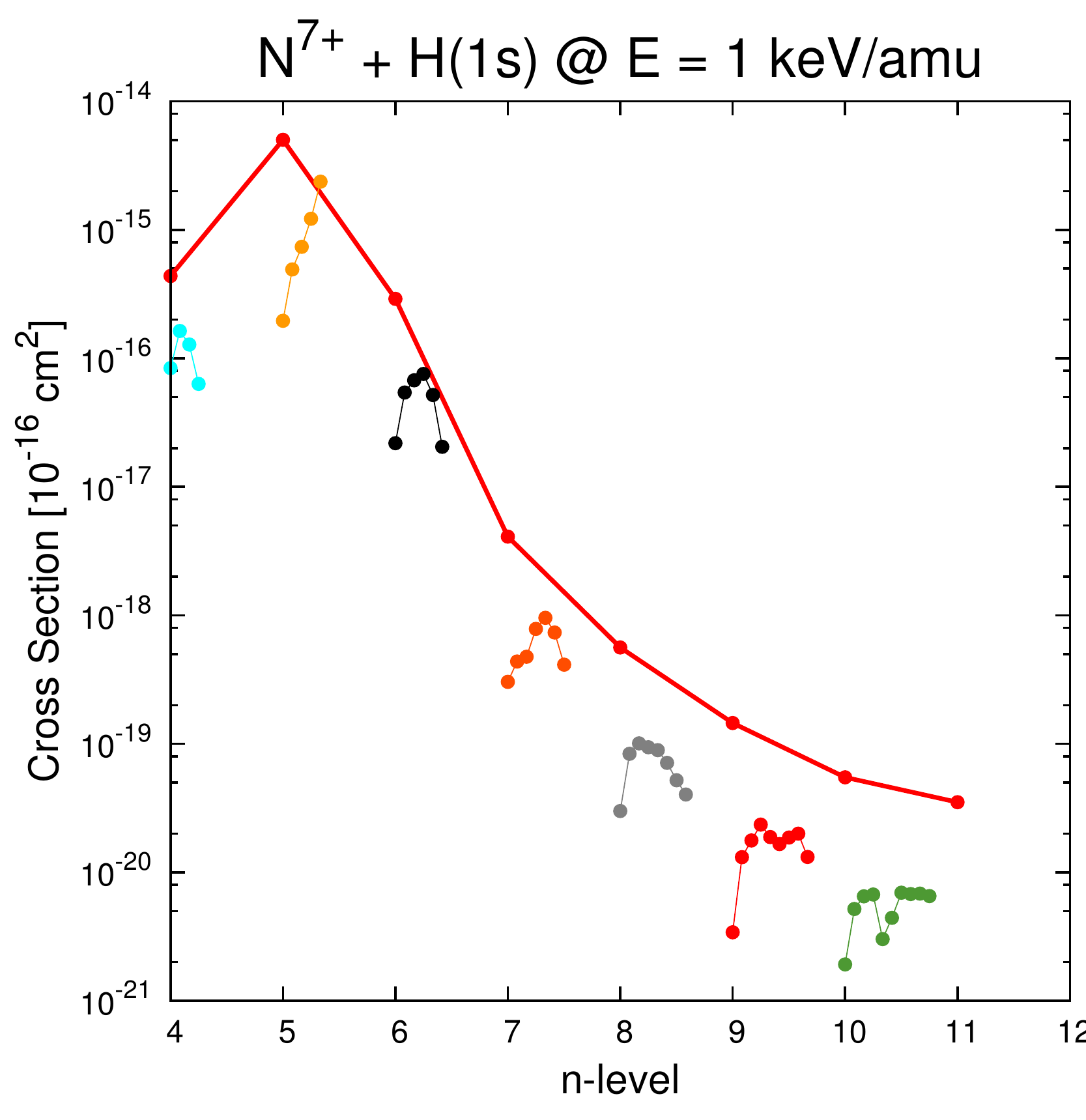}}
\subfigure[\label{f:n7hnl_1-b}]{\includegraphics[width=.49\columnwidth]{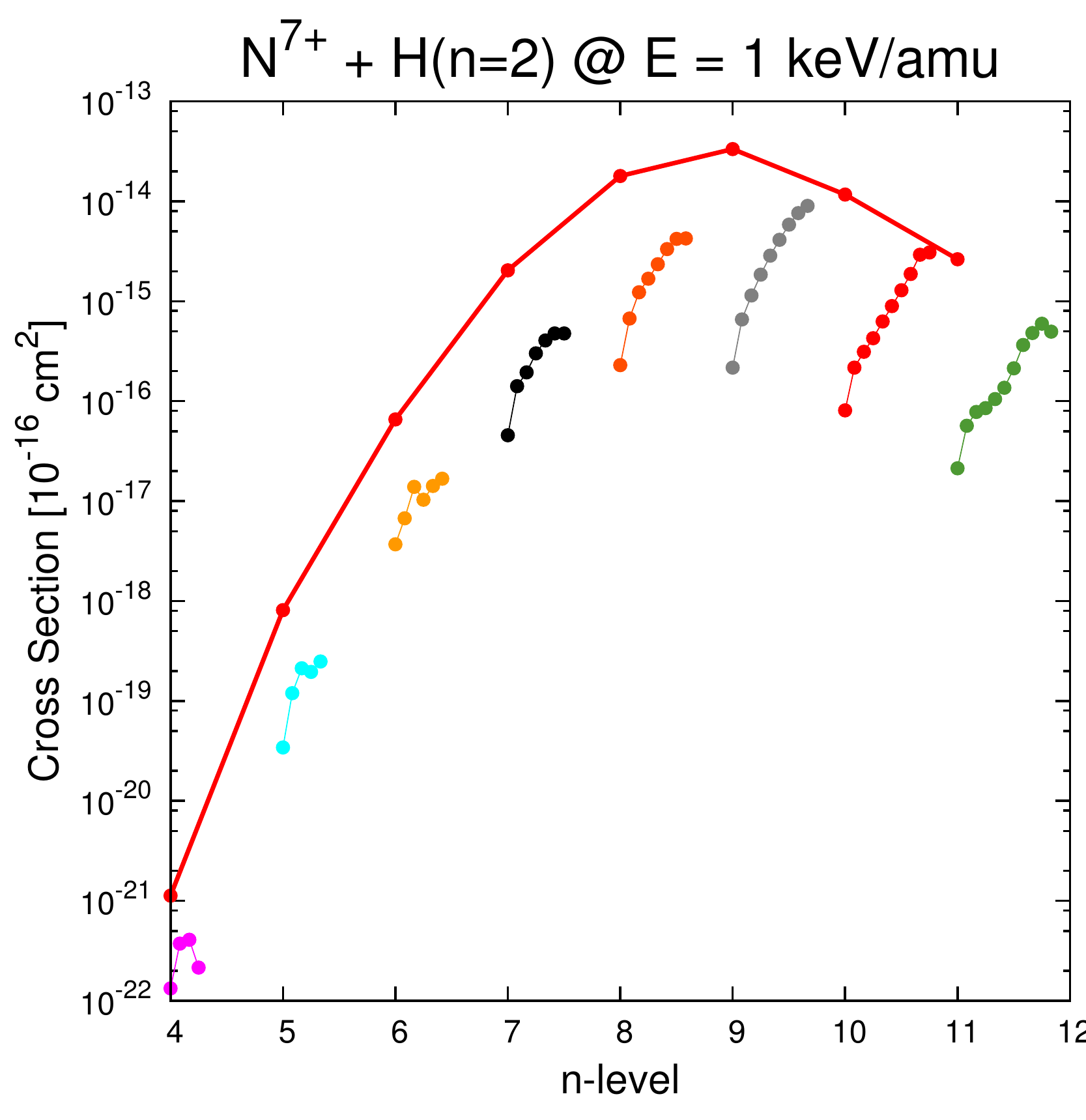}}
\caption[$n\ell$-resolved cross sections for N$^{7+}$ + H($n=1,2$) collisions at E = 1 keV/amu]{$n\ell$-resolved cross sections for N$^{7+}$ impact on (a) H(1s) and (b) H($n=2$) collisions at E = 1 keV/amu. The red data on top show the $n$-resolved cross sections as a function of the principal quantum number $n$. The differently colored lines below the red line show the $n\ell$-resolved cross sections for the $n$ quantum number indicated on the x-axis, each starting with $\ell = 0$ on the left and ending with $\ell = n-1$ on the right.}
\label{f:n7hnl_1}
\end{figure}

\begin{figure}
\centering
\subfigure[\label{f:n7hnl_10-a}]{\includegraphics[width=.49\columnwidth]{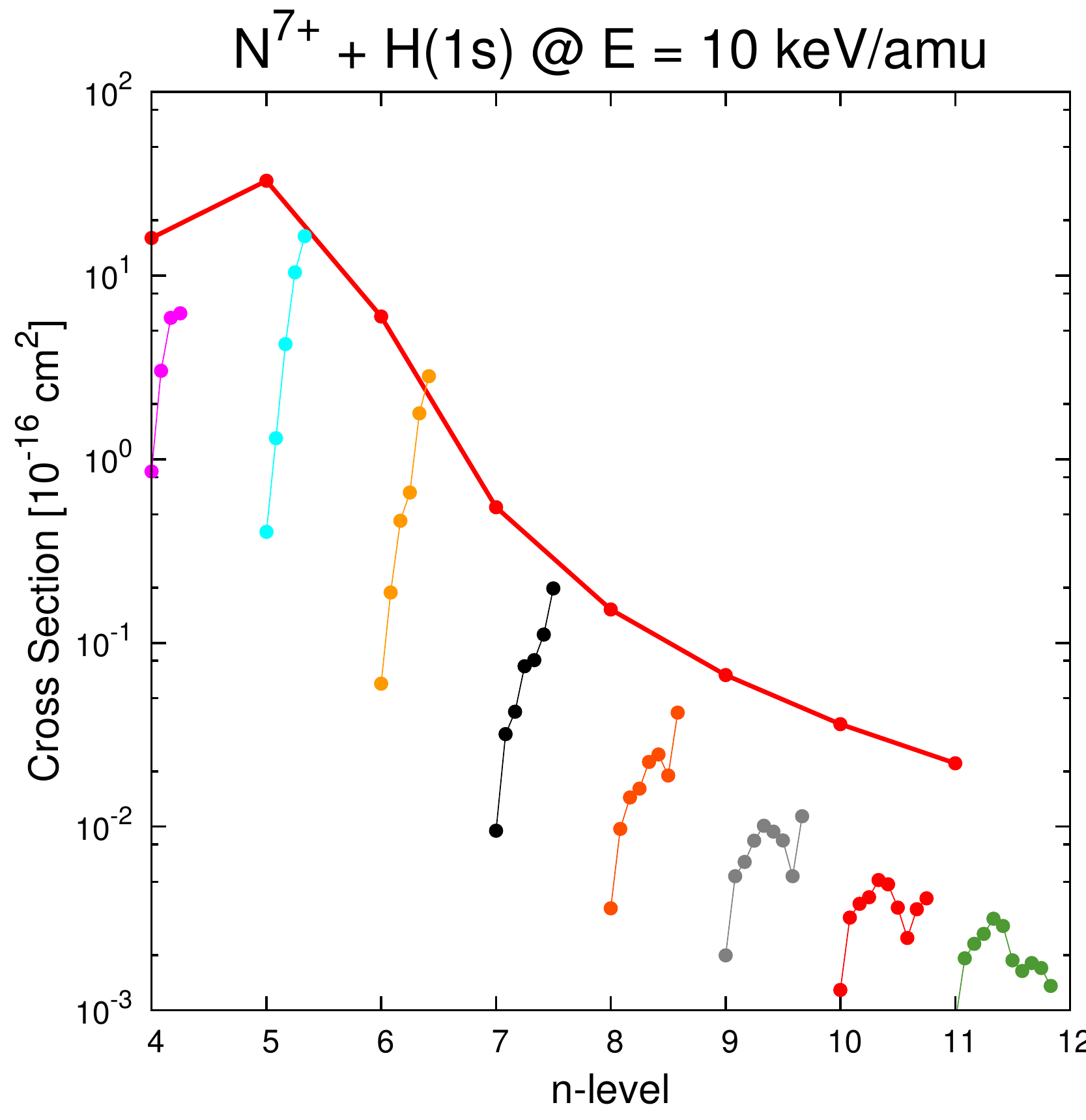}}
\subfigure[\label{f:n7hnl_10-b}]{\includegraphics[width=.49\columnwidth]{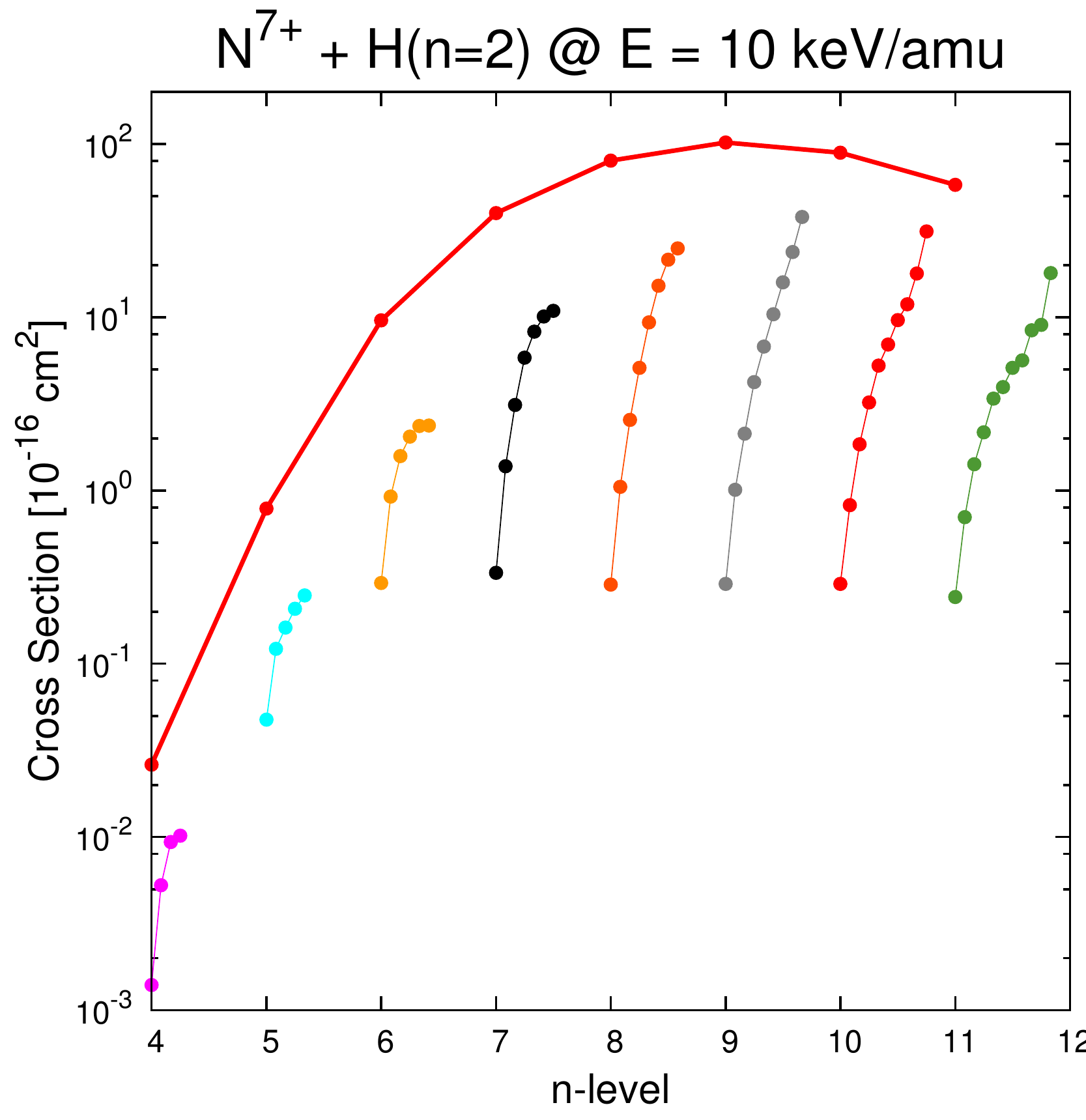}}
\caption[$n\ell$-resolved cross sections for N$^{7+}$ + H($n=1,2$) collisions at E = 10 keV/amu]{The same as fig.\ref{f:n7hnl_1} but for E = 10 keV/amu.}
\label{f:n7hnl_10}
\end{figure}

\begin{figure}
\centering
\subfigure[\label{f:n7hnl_100-a}]{\includegraphics[width=.49\columnwidth]{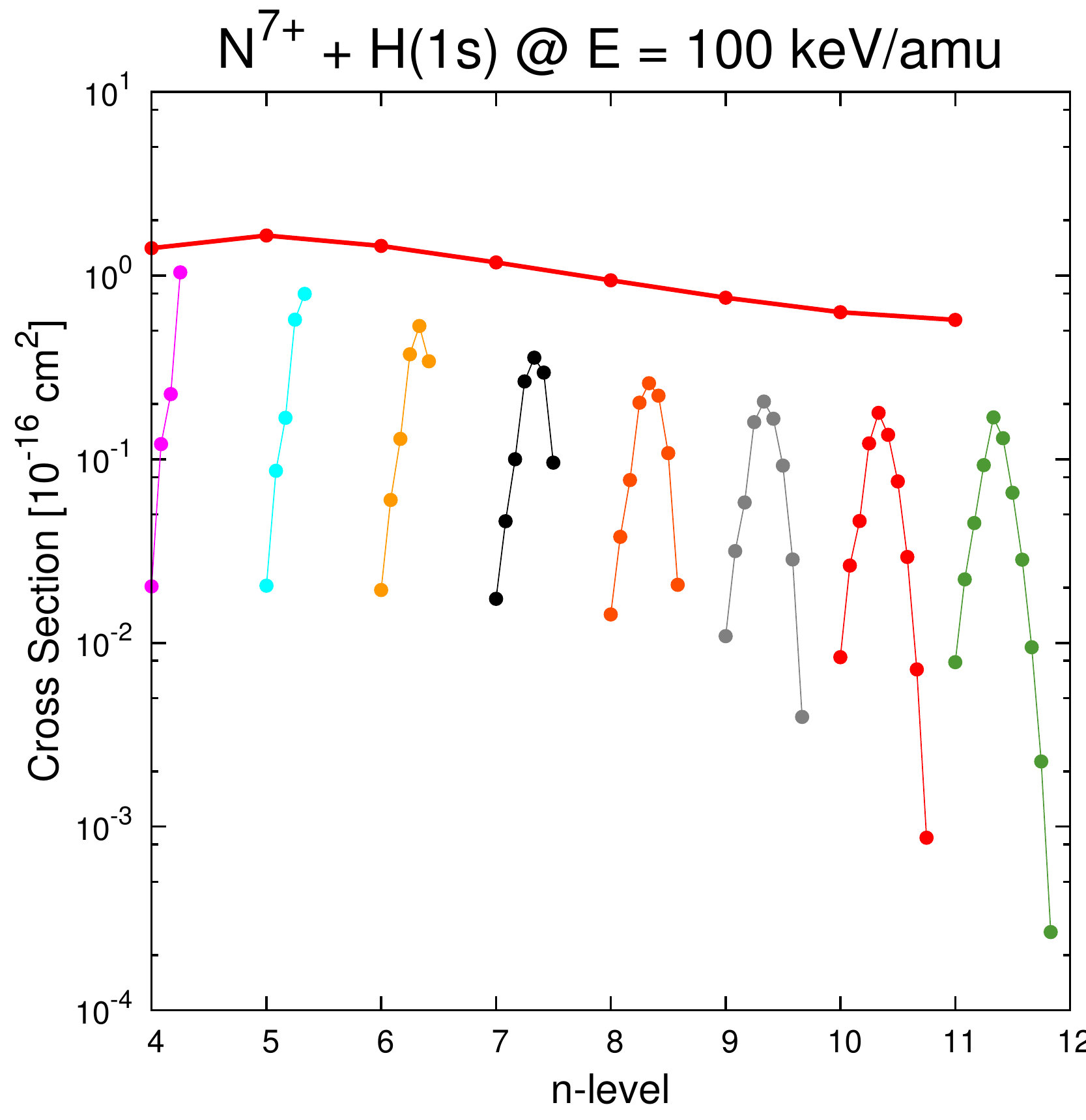}}
\subfigure[\label{f:n7hnl_100-b}]{\includegraphics[width=.49\columnwidth]{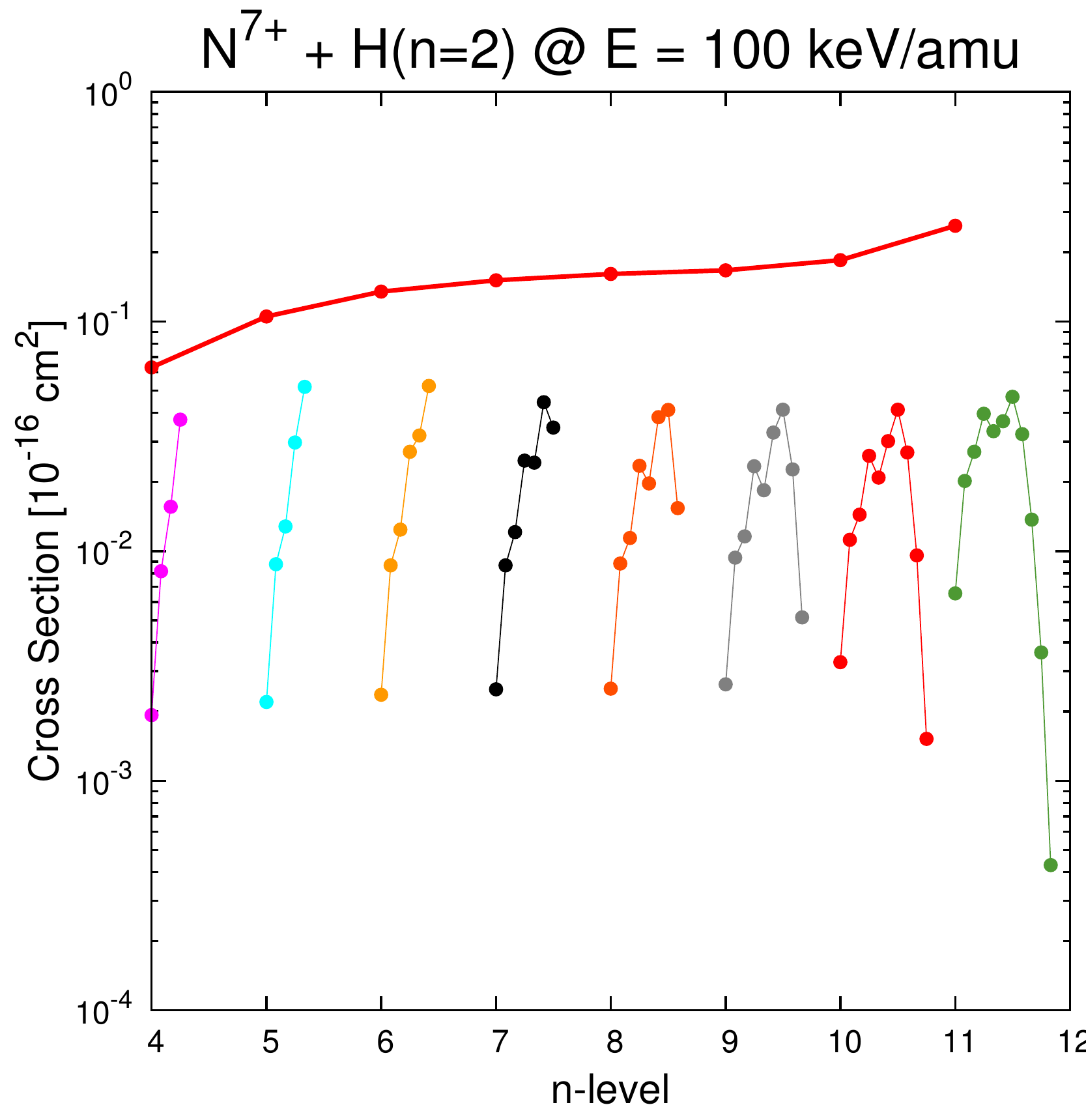}}
\caption[$n\ell$-resolved cross sections for N$^{7+}$ + H($n=1,2$) collisions at E = 10 keV/amu]{The same as fig.\ref{f:n7hnl_1} but for E = 100 keV/amu.}
\label{f:n7hnl_100}
\end{figure}

The wavelengths of the respective deexcitation lines, $n_{i} \rightarrow n_{f}$, can be calculated using Rydberg's formula.  The wavelength of a couple of transitions (8 $\rightarrow$ 7, 9 $\rightarrow$ 8 and  10 $\rightarrow$ 8) are in the visible range. Figures \ref{f:n7hnl_1}, \ref{f:n7hnl_10} and \ref{f:n7hnl_100} show $n\ell$-resolved cross sections for E = 1, 10 and 100 keV/amu. The red curves show $n$-resolved cross sections as a function of the principal quantum number $n$. For ground state hydrogen targets the main capture channel at $n=5$ is very pronounced at E = 1 keV/amu and E = 10 keV/amu, whereas  at E = 100 keV/amu the distribution is flatter, but $n=5$ is still the largest partial cross section. For excited state targets the classical over the barrier model~\cite{ryufuku80} yields $n=9$ as strongest populated shell. This can be seen in figures \ref{f:n7hnl_1-b} \& \ref{f:n7hnl_10-b}. But the results for E = 100 keV/amu, fig. \ref{f:n7hnl_100-b}, show a different behavior. A similar trend can be observed in Be$^{4+}$ + H($n=2$) collisions where at impact energies of about 100 keV/amu the largest included $n$-shell shows the largest capture cross sections \cite{igenbergs09}. Above $\approx$ 200 keV/amu the general trend of the $n$-resolved cross sections follows the 1/$n^3$ rule \cite{coleman68}.

%%%%%%%%%%%%%%%%%%%%%%%%%%%%%%%%%%%%%%%%%%%%%%%%%%%%%%%
\section{Comparison of $N^{6+}$ - H and C$^{6+}$+H}

For heavier impurities, e.g., noble gas ions like Ne or Ar, the fractional abundances of ions species other than fully stripped at typical plasma temperatures are non-negligible. In single-electron transfer collisions, the AOCC method treats all passive electrons as perturbations to the potential of the active electron in the respective collision centre. This means that elaborate potentials need to be found resulting in much more complex structures of the matrix elements. It is, nevertheless, a reasonable assumption that the influence of closely bound core electrons on the active electron that captures into very high $n$-shells is negligible. We therefore conducted a study of C$^{6+}$ + H($n=1,2$) in comparison to N$^{6+}$ + H($n=1,2$) to see whether this assumption is valid. For H(1s) the main capture channel of the active electron is $n=4$, which is of course much lower than in the case of Ne or Ar. One would therefore expect the difference in the cross sections as a result of the perturbed potential to be more pronounced. 

To determine the appropriate pseudopotential we used the Hartree-Fock method in the frozen-core approximation~\cite{froese97}:

\begin{equation}
V^{N^{6+}} (r) = - \frac {6}{r}-\frac{e^{-14r}}{r}- 7 e^{-14r}
\label{e:pspot}
\end{equation}

\begin{figure}
\centering
\includegraphics[width=7.cm]{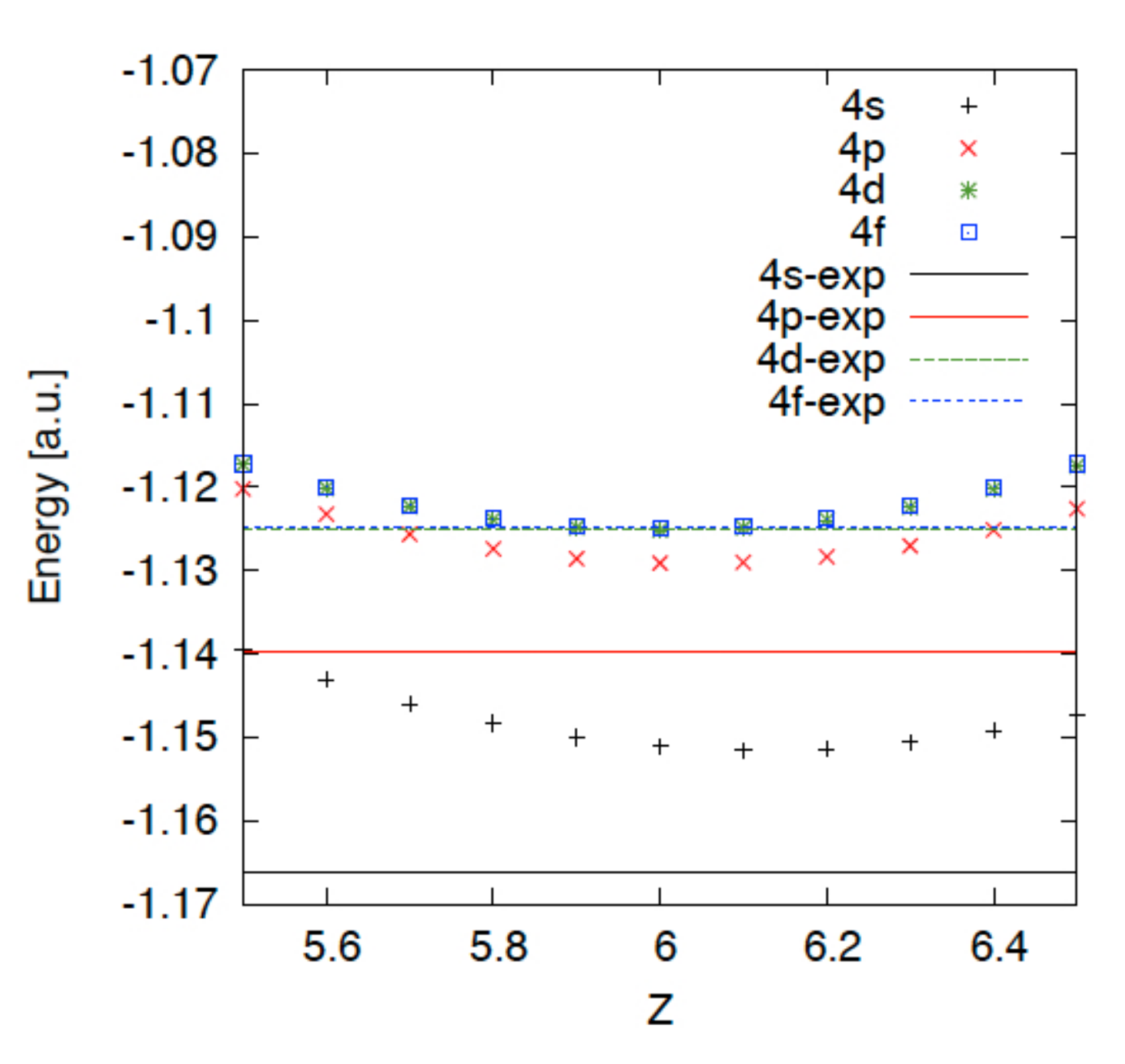}
\caption{(Color online) Eigenenergies of the bound states of the active electron in the frozen-core potential of N$^{6+}$, see (\ref{e:pspot}) in comparison with experimental values~\protect\cite{bashkin78}.}
\label{f:N6Zscan}
\end{figure}

Subsequently, the appropriate basis states on the N$^{6+}$ centre had to be determined. Since we use basis functions with Laguerre-type radial parts we can vary $Z$ to approximate the basis functions on the N$^{6+}$ centre. Fig. \ref{f:N6Zscan} shows the eigenenergies for several possible $Z$ in comparison with spectroscopic data from~\cite{bashkin78}. The minimum values occur at $Z=6.1$.

\begin{figure}
\centering
\subfigure[\label{f:c6n6tot-a}]{\includegraphics[width=.48\columnwidth]{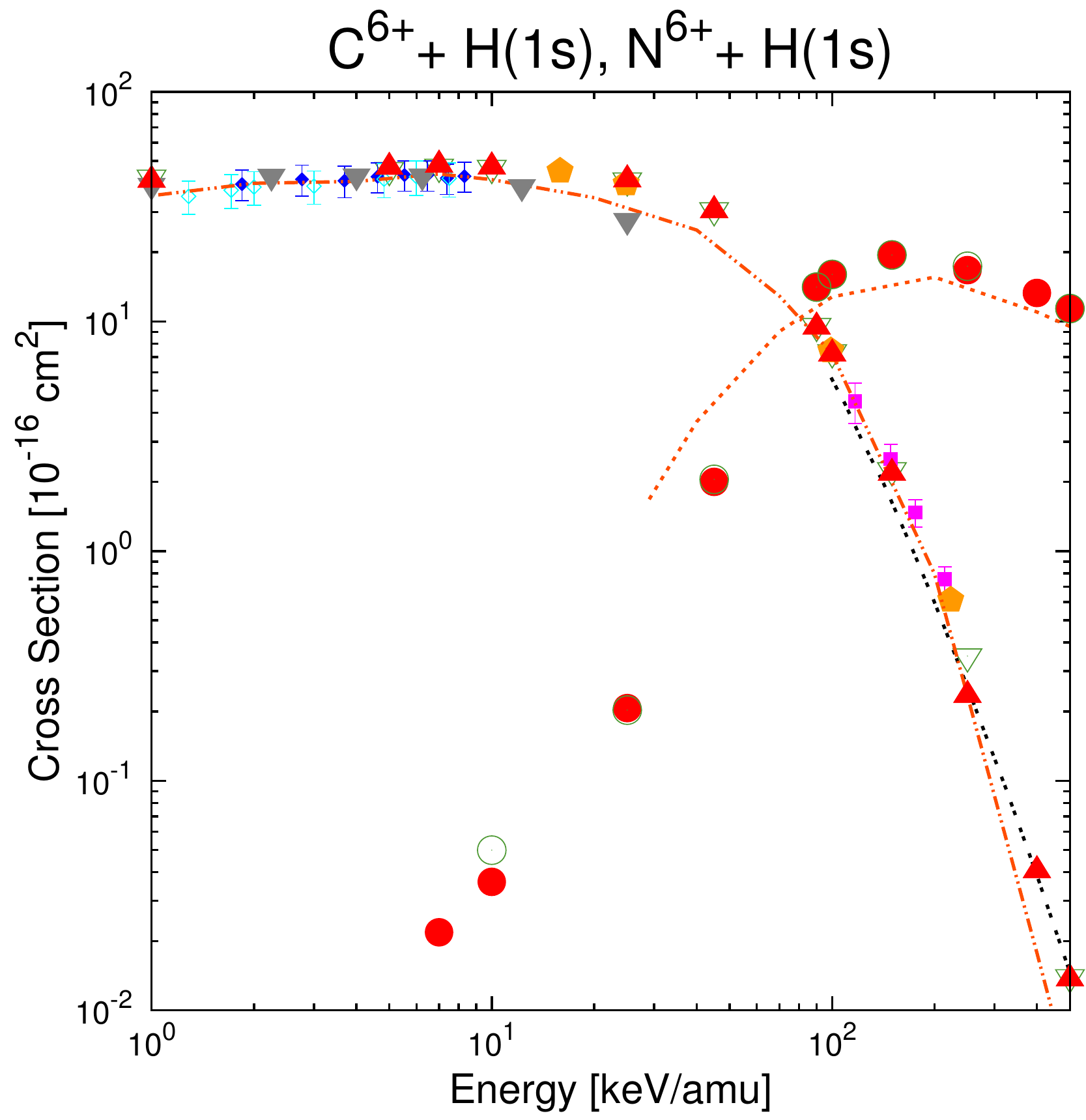}}
\subfigure[\label{f:c6n6tot-b}]{\includegraphics[width=.48\columnwidth]{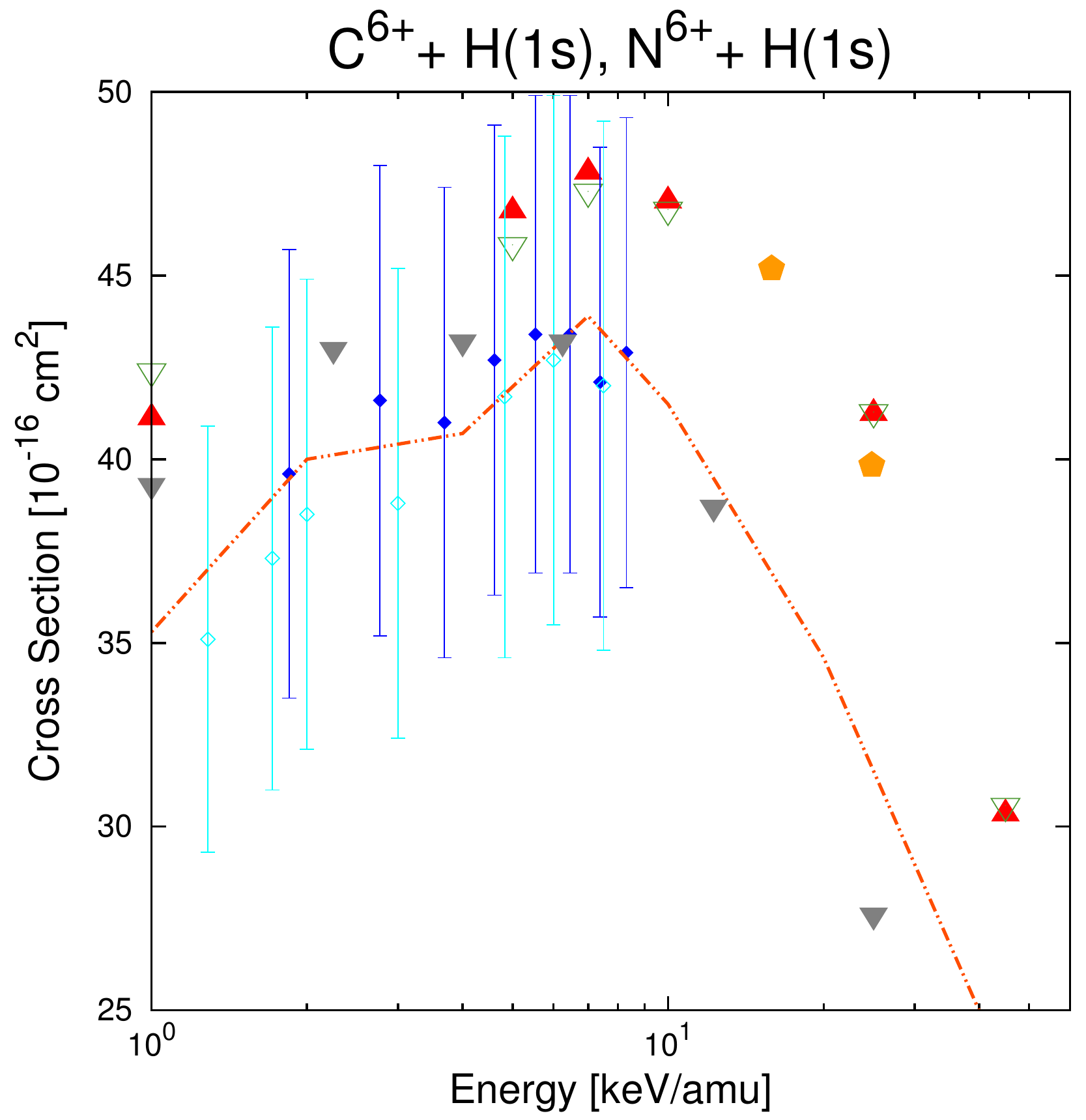}}
\caption{(Color online) Comparison between C$^{6+}$ + H(1s) (full, red symbols) and N$^{6+}$ + H(1s) (open, green symbols). AOCC data (CX \protect\gplps{9}\protect\gplps{10}, ION \protect\gplps{7}\protect\gplps{6}) are compared to experimental data~\protect\cite{meyer85} \mbox{(\protect\gplps{13}\protect\gplps{12})} and~\protect\cite{goffe79} (\protect\gplps{5}), CTMC data~\protect\cite{illescas99} (\protect\gplps{15}), CDW data~\protect\cite{busnengo97} (\protect\gplls{7}), recommended cross sections~\protect\cite{janev88} (CX\protect\gplls{8}, ION\protect\gplls{3}) and AO+ data~\protect\cite{fritsch84} (\protect\gplps{11}). (a) Full energy range. (b) Energy range between 1-60 keV/amu using an expanded linear scale on the ordinate.}
\label{f:c6n61stot}
\end{figure}

\begin{figure}
\centering
\includegraphics[width=.48\columnwidth]{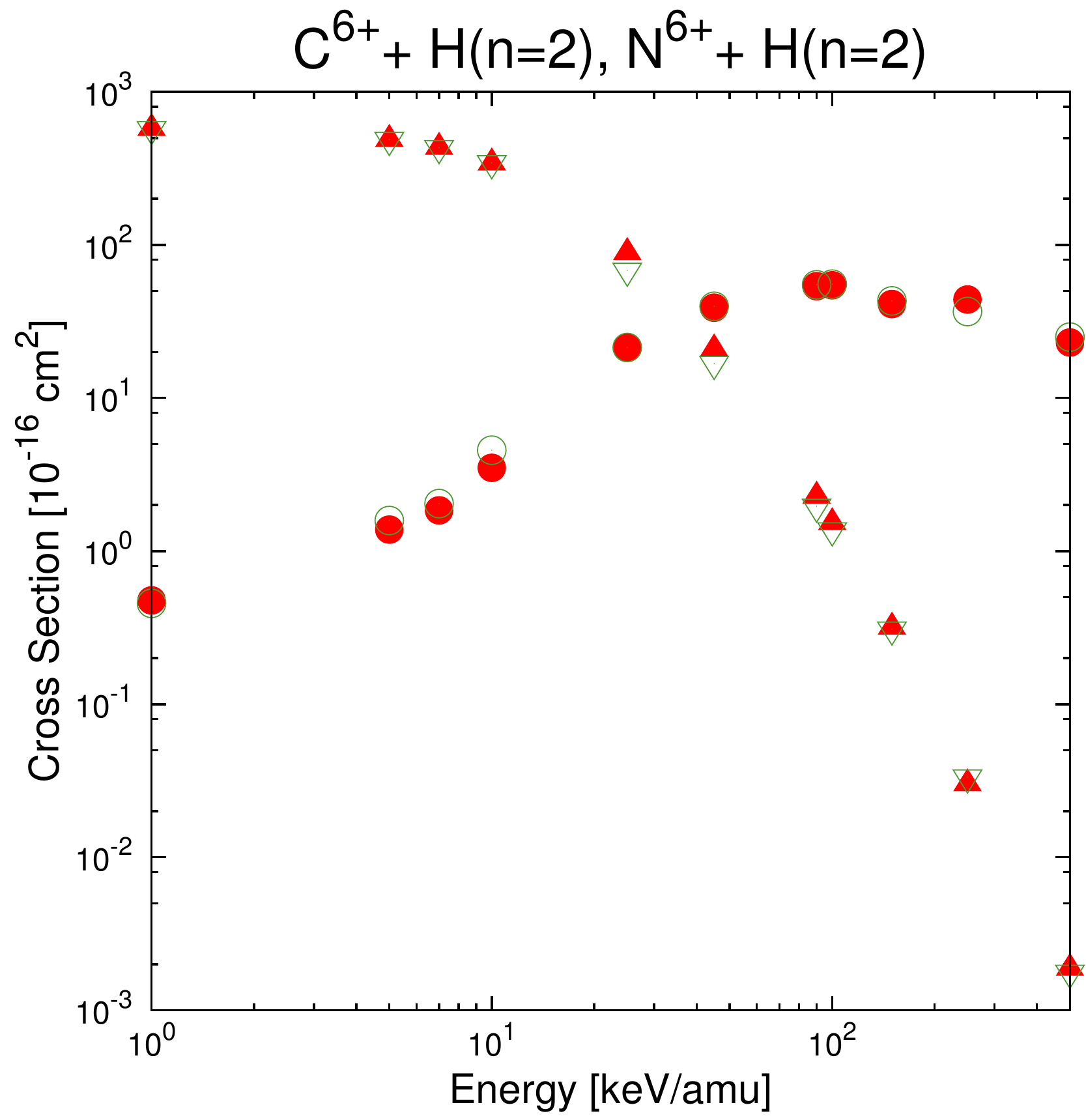}
\caption{(Color online) Same as fig.\ref{f:c6n61stot} but for an H($n=2$) target. \mbox{C$^{6+}$ + H($n=2$)} (full, red symbols) total CX (\protect\gplps{9}) and ION (\protect\gplps{7}) cross sections are compared those for \mbox{N$^{6+}$ + H($n=2$)} (open, green symbols) for CX (\protect\gplps{10}) and ION (\protect\gplps{6}).}
\label{f:c6n6n2tot}
\end{figure}

Figs.\ref{f:c6n61stot}\hspace*{2pt}\&\hspace*{2pt}\ref{f:c6n6n2tot} show total CX and ION cross sections for both targets (ground and excited state H) and both projectiles (C$^{6+}$, N$^{6+}$). The data for the H(1s) targets are compared to experimental data for both systems from~\citeasnoun{meyer85} and in addition for C$^{6+}$ +  H(1s) from~\citeasnoun{goffe79}, as well as AO+ data from~\citeasnoun{fritsch84} and CTMC cross sections from~\citeasnoun{illescas99}. At low energies, well below 10 keV, there is a certain difference, but this paper focusses at fusion relevant energy regions. Typical injector energies for hydrogen/deuterium heating or diagnostic beams are in the order of a few 10 keV/amu to 100 keV/amu. Here it becomes obvious that the cross sections for the different impurity ions do not deviate at all anymore. 

\begin{figure}
\centering
\subfigure[\label{f:c6n6n-a}]{\includegraphics[width=.48\columnwidth]{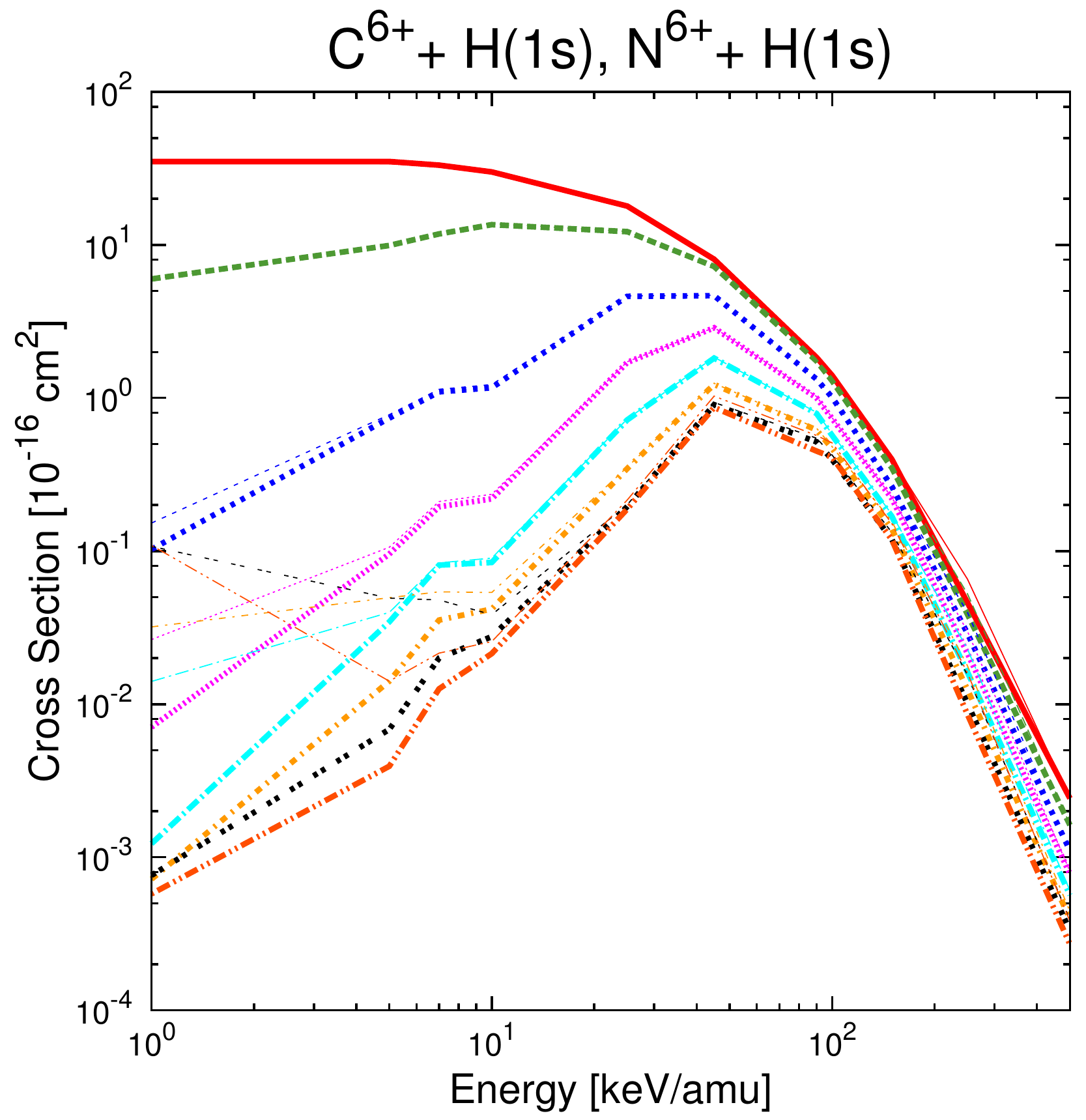}}
\subfigure[\label{f:c6n6n-b}]{\includegraphics[width=.48\columnwidth]{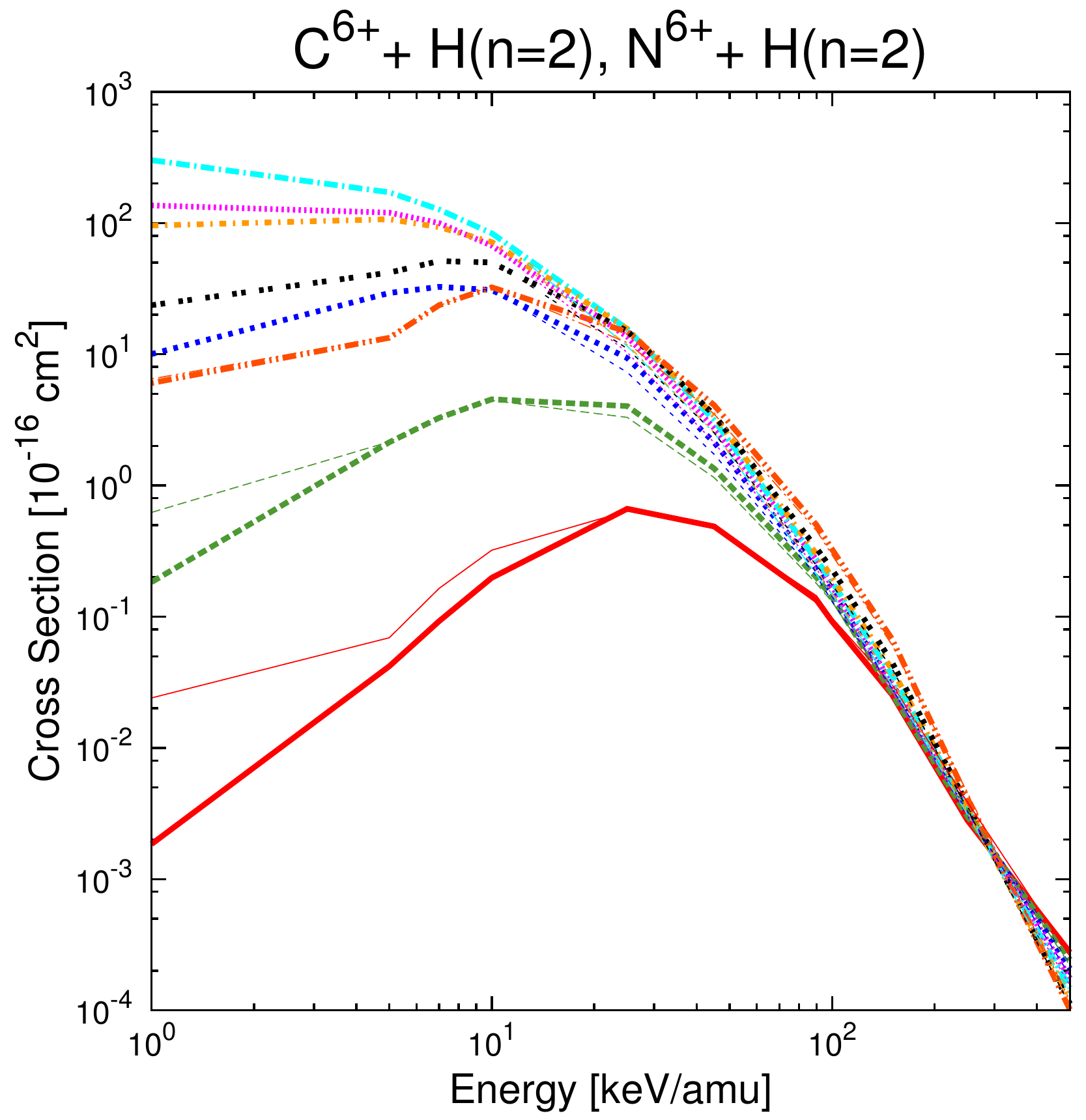}}
\caption{(Color online) $n$-resolved CX cross sections from AOCC calculations for C$^{6+}$ (thick lines) and N$^{6+}$ (thin lines) projectiles impacting on (a) H(1s). (b) H($n=2$). In both plots $\sigma_n$ are shown in the same colors and linestyles: $n=4$ (red,\protect\gplls{1}), $n=5$ (green,\protect\gplls{2}), $n=6$ (blue,\protect\gplls{3}), $n=7$ (magenta,\protect\gplls{4}), $n=8$ (cyan,\protect\gplls{5}), $n=9$ \mbox{(light orange,\protect\gplls{6})}, $n=10$ (black,\protect\gplls{7}), and $n=11$ (dark orange,\protect\gplls{8}).}
\label{f:c6n6n}
\end{figure}

The same behaviour can be observed in fig.\ref{f:c6n6n} which shows $n$-resolved capture cross sections, i.e. deviations only occur at low impact energies. It can be seen that the main capture channel is the $n=4$ shell for H(1s) targets and the $n=8$ shell for H($n=2$) targets. In both cases large deviations only occur for those cross sections that are already very small. Also, only subshell cross sections for shells with $n\geq 8$ deviate significantly for ground state H targets. C$^{5+} (n=8)$ is the main capture channel for charge exchange from excited state H($n=2$). The increase of the subshell cross sections at low energies for $n \geq 8$ can thus be explained by an intermediate excitation to H($n=2$) during the collision and then subsequent charge exchange from the excited state. It can be shown that removing all excited states from the basis also removes this effect. \cite{igenbergsPhD}

Thus, we can conclude that the inclusion of a closely bound passive electron on the ion neither influences  the total nor the $n$-resolved cross sections. All AOCC results, including $n\ell$-resolved cross sections, are summarized in Tables \ref{t:n6h1s} to \ref{t:c6hn2}.

%%%%%%%%%%%%%%%%%%%%%%%%%%%%%%%%%%%%%%%%%%%%%%%%%%%%%%%
\section{Conclusions}

We presented total and state-resolved charge exchange and ionisation cross sections for three different ions colliding with ground and excited state neutral hydrogen.

In a first part, we focussed on fully stripped nitrogen ions, N$^{7+}$. For impact on ground state H(1s), our AOCC calculations for total charge exchange and ionisation agree excellently with data from the literature. Additionally, we discussed $n$- and $n\ell$- resolved CX cross sections that are urgently needed for the evaluation of spectroscopic data from CXRS of fusion plasmas.

Excited state H($n=2$) targets are important for the analysis of diagnostic data because a small, but non-negligible fraction of the neutral heating beam at mid-sized tokamaks like ASDEX-Upgrade or JET is in this state. Charge exchange and ionisation cross sections are generally approximately an order of magnitude larger than for H(1s). We complemented our AOCC calculations with additional CTMC calculations to make up for the absence of reference data from the literature. It is worth noting that these CTMC calculation coincide excellently with the scaling method ADAS315 (which uses CTMC results from different collision systems). Actual evaluations of spectroscopic data will have to show which trend is likely to be better. Nevertheless the differences in the density profiles evaluated on the basis of our AOCC calculations for H($n=2$) or on the basis of our CTMC calculations for H($n=2$) might be to small to tell which data set makes more sense. The use of velocity averaged effective emission cross sections and the relatively small percentage of excited state H in the beam reduce the sensitivity of the evaluated results to the actual CX cross sections from excited state hydrogen.

In the second part, we compared two collision systems with equally charged ions, namely C$^{6+}$ + H($n=1,2$) and N$^{6+}$ + H($n=1,2$). In the latter case we modeled the influence of the passive electron with a pseudopotential. The resulting cross sections show excellent agreement with both experiment and other theoretical approaches. Furthermore little to no deviation can be observed between the two projectiles. Therefore we can conclude that the role of the passive electron is nearly negligible, especially in the energy ranges of interest for nuclear fusion research.

%%%%%%%%%%%%%%%%%%%%%%%%%%%%%%%%%%%%%%%%%%%%%%%%%%%%%%%
\section{Acknowledgments}

Katharina Igenbergs gratefully acknowledges the support received through a grant from the Friedrich Schiedel Foundation for Energy Technology.

The computational results presented in this paper have been achieved in part using the Vienna Scientific Cluster (VSC) and High Performance Computing For Fusion (HPC-FF).

This work, supported by the European Commission under the Contract of Association between EURATOM and \"OAW, was carried out within the framework of the European Fusion Development Agreement (EFDA). The views and opinions expressed
herein do not necessarily reflect those of the European Commission.

%%%%%%%%%%%%%%%%%%%%%%%%%%%%%%%%%%%%%%%%%%%%%%%%%%%%%%%
\section{Data Tables}

\begin{table}
 \caption{Data for N$^{7+}$ + H(1s)}
 \vspace*{2pt}
\tiny
\centering
\tabcolsep=0.11cm
\begin{tabular}{S{p{2pt}} S{p{2pt}} Sc Sc Sc Sc Sc Sc Sc Sc Sc Sc Sc Sc}
\toprule
 & & \multicolumn{10}{c}{\bf \scriptsize impact energies $[$kev/amu$]$} & & \\
& & \scriptsize 1.0 & \scriptsize 5.0 & \scriptsize 7.0 & \scriptsize 10.0 & \scriptsize 25.0 & \scriptsize 30.0 & \scriptsize 45.0 & \scriptsize90.0 & \scriptsize 100.0 & \scriptsize 150.0 & \scriptsize 250.0 & \scriptsize 500.0 \\
\otoprule
 & \multicolumn{12}{c}{\bf \scriptsize total cross sections [$\bf10^{-16}$ cm$\bf ^2$]} & \\
 \multicolumn{2}{Sc}{CX} & 5.74E+01 & 5.74E+01 & 5.78E+01 & 5.58E+01 & 4.81E+01 & 4.54E+01 & 3.60E+01 & 1.18E+01 & 9.13E+00 & 2.68E+00 & 4.48E-01 & 1.41E-02 \\
 \multicolumn{2}{Sc}{ION} &  8.21E-04 & 1.41E-02 & 4.02E-02 & 5.67E-02 & 3.71E-01 & 7.95E-01 & 4.35E+00 & 2.18E+01 & 2.34E+01 & 2.44E+01 & 2.15E+01 & 1.44E+01\\
\otoprule
 & \multicolumn{12}{Sc}{\bf \scriptsize state resolved CX cross sections [$\bf10^{-16}$ cm$\bf ^2$]} & \\
\scriptsize $n$ & \scriptsize $\ell$ & & & & & & & & & & \\
\midrule
  2 & & 5.86E-10& 3.10E-09& 9.70E-08& 1.97E-06& 2.73E-04& 5.53E-04& 1.94E-03& 6.41E-03& 6.78E-03& 1.05E-02& 6.40E-03& 9.00E-04 \\
\bottomrule
 \end{tabular}
 \label{t:n7h1s}
 \end{table}
 
 \begin{table}
 \caption{Data for N$^{7+}$ + H($n=2$)}
 \vspace*{2pt}
\tiny
\centering
\tabcolsep=0.11cm
\begin{tabular}{S{p{2pt}} S{p{2pt}} Sc Sc Sc Sc Sc Sc Sc Sc Sc Sc Sc Sc}
\toprule
 & & \multicolumn{10}{c}{\bf \scriptsize impact energies [kev/amu]} & & \\
& & \scriptsize 1.0 & \scriptsize 5.0 & \scriptsize 7.0 & \scriptsize 10.0 & \scriptsize 25.0 & \scriptsize 30.0 &\scriptsize 45.0 & \scriptsize90.0 & \scriptsize 100.0 & \scriptsize 150.0 & \scriptsize 250.0 & \scriptsize 500.0 \\
\otoprule
 & \multicolumn{12}{c}{\bf \scriptsize total cross sections [$\bf10^{-16}$ cm$\bf ^2$]} & \\
 \multicolumn{2}{Sc}{CX} & 6.77E+02 & 5.52E+02 & 4.82E+02 & 3.80E+02 & 8.22E+01 & 5.05E+01 & 1.63E+01 & 1.92E+00 & 1.25E+00 & 2.54E-01 & 2.40E-02 & 4.60E-04\\
 \multicolumn{2}{Sc}{ION} & 4.63E+00 & 3.97E+00 & 8.54E+00 & 2.67E+01 & 1.75E+02 & 2.06E+02 & 2.35E+02 & 2.38E+02 & 2.32E+02 & 1.99E+02 & 1.51E+02 & 9.26E+01 \\
\otoprule
 & \multicolumn{12}{Sc}{\bf \scriptsize state resolved CX cross sections [$\bf10^{-16}$ cm$\bf ^2$]} & \\
\scriptsize $n$ & \scriptsize $\ell$ & & & & & & & & & & \\
\midrule
  2 & & 1.18E-07& 5.46E-08& 6.83E-08& 3.57E-08& 1.06E-06& 3.08E-06& 2.78E-05& 1.62E-04& 2.06E-04& 2.36E-04& 1.62E-04& 2.11E-05 \\
\bottomrule
 \end{tabular}
 \label{t:n7hn2}
 \end{table}
 
 \begin{table}
 \caption{Data for N$^{6+}$ + H(1s)}
 \vspace*{2pt}
\tiny
\centering
\begin{tabular}{S{p{2pt}} S{p{2pt}} Sc Sc Sc Sc Sc Sc Sc Sc Sc Sc Sc}
\toprule
 & & \multicolumn{11}{c}{\bf \scriptsize impact energies $[$kev/amu$]$} \\
 & & \scriptsize 1.0 & \scriptsize 5.0 & \scriptsize 7.0 & \scriptsize 10.0 & \scriptsize 25.0 & \scriptsize 45.0 & \scriptsize 90. & \scriptsize 100. & \scriptsize 150. & \scriptsize 250. & \scriptsize 500. \\
\otoprule
 & & \multicolumn{11}{c}{\bf \scriptsize total cross sections [$\bf10^{-16}$ cm$\bf ^2$]}\\
 \multicolumn{2}{Sc}{CX} & 4.24E+1 & 4.58E+1 & 4.73E+1 & 4.68E+1 & 4.13E+1 & 3.06E+1 & 9.60E+0 & 7.37E+0 & 2.26E+0 & 3.49E-1 & 1.40E-2 \\
 \multicolumn{2}{Sc}{ION} & 6.37E-2 & 1.09E-1 & 8.18E-2 & 4.98E-2 & 2.03E-1 & 2.05E+0 & 1.41E+1 & 1.60E+1 & 1.95E+1 & 1.74E+1 & 1.14E+1\\
\otoprule
 & & \multicolumn{11}{Sc}{\bf \scriptsize state resolved CX cross sections [$\bf10^{-16}$ cm$\bf ^2$]}\\
\scriptsize $n$ & \scriptsize $\ell$ & \multicolumn{11}{c}{} \\
\midrule
2 & & 6.00E-7 & 3.11E-6 & 1.61E-5 & 2.06E-4 & 8.00E-3 & 2.81E-2 & 5.06E-2 & 5.15E-2 & 5.84E-2 &  2.30E-2 & 2.79E-3  \\
\bottomrule
 \end{tabular}
 \label{t:n6h1s}
 \end{table}
 
 \begin{table}
 \caption{Data for N$^{6+}$ + H($n=2$)}
 \vspace*{2pt}
\tiny
\centering
\tabcolsep=0.11cm
\begin{tabular}{S{p{2pt}} S{p{2pt}} Sc Sc Sc Sc Sc Sc Sc Sc Sc Sc Sc}
\toprule
 & & \multicolumn{11}{c}{\bf \scriptsize impact energies $[$kev/amu$]$} \\
 & & \scriptsize 1.0 & \scriptsize 5.0 & \scriptsize 7.0 & \scriptsize 10.0 & \scriptsize 25.0 & \scriptsize 45.0 & \scriptsize 90. & \scriptsize 100. & \scriptsize 150. & \scriptsize 250. & \scriptsize 500. \\
\otoprule
 & & \multicolumn{11}{c}{\bf \scriptsize total cross sections [$\bf10^{-16}$ cm$\bf ^2$]}\\
 \multicolumn{2}{Sc}{CX} &  5.75E+02 & 4.88E+02 & 4.32E+02 & 3.44E+02 & 8.99E+01 & 2.16E+01 & 2.38E+00 & 1.63E+00 & 3.50E-01 & 3.57E-02 & 1.81E-03\\
  \multicolumn{2}{Sc}{ION} & 4.55E-01 & 1.58E+00 & 2.05E+00 & 4.57E+00 & 2.48E+01 & 5.17E+01 & 7.46E+01 & 7.51E+01 & 5.89E+01 & 4.85E+01 & 3.19E+01\\
\otoprule
 & & \multicolumn{11}{Sc}{\bf \scriptsize state resolved CX cross sections [$\bf10^{-16}$ cm$\bf ^2$]}\\
\scriptsize $n$ & \scriptsize $\ell$ & \multicolumn{11}{c}{} \\
  2 & & 8.78E-05& 1.39E-03& 2.45E-03& 4.36E-03& 1.28E-02& 4.01E-02& 3.20E-02& 3.43E-02& 1.18E-02& 1.31E-03& 1.96E-04 \\
\bottomrule
 \end{tabular}
  \label{t:n6hn2}
 \end{table}

\begin{table}
 \caption{Data for C$^{6+}$ + H(1s)}
 \vspace*{2pt}
\tiny
\centering
\begin{tabular}{S{p{2pt}} S{p{2pt}} Sc Sc Sc Sc Sc Sc Sc Sc Sc Sc Sc}
\toprule
 & & \multicolumn{11}{c}{\bf \scriptsize impact energies $[$kev/amu$]$} \\
 & & \scriptsize 1.0 & \scriptsize 5.0 & \scriptsize 7.0 & \scriptsize 10.0 & \scriptsize 25.0 & \scriptsize 45.0 & \scriptsize 90. & \scriptsize 100. & \scriptsize 150. & \scriptsize 250. & \scriptsize 500. \\
\otoprule
 & & \multicolumn{11}{c}{\bf \scriptsize total cross sections [$\bf10^{-16}$ cm$\bf ^2$]}\\
 \multicolumn{2}{Sc}{CX} & 4.11E+1 & 4.67E+1  & 4.78E+1 & 4.70E+1 & 4.12E+1 & 3.03E+1 & 9.42E+0 & 7.20E+0 & 2.18E+0 & 2.34E-1 & 1.37E-2 \\
 \multicolumn{2}{Sc}{ION} & 3.29E-4 & 8.18E-3 & 2.19E-2 & 4.98E-02 & 2.03E-01 & 2.05E+00 & 1.41E+01 & 1.60E+01 & 1.95E+01 & 1.74E+01 & 1.14E+01 \\
\otoprule
 & & \multicolumn{11}{Sc}{\bf \scriptsize state resolved CX cross sections [$\bf10^{-16}$ cm$\bf ^2$]} \\
\scriptsize $n$ & \scriptsize $\ell$ & \multicolumn{11}{c}{} \\
\midrule
  2 & & 3.03E-06& 2.19E-06& 1.88E-05& 1.68E-04& 9.79E-03& 3.33E-02& 5.41E-02& 5.88E-02& 5.83E-02& 1.04E-02& 2.79E-03 \\
\bottomrule
 \end{tabular}
 \label{t:c6h1s}
 \end{table}
 
 \begin{table}
 \caption{Data for C$^{6+}$ + H($n=2$)}
 \vspace*{2pt}
\tiny
\centering
\begin{tabular}{S{p{2pt}} S{p{2pt}} Sc Sc Sc Sc Sc Sc Sc Sc Sc Sc Sc}
\toprule
 & & \multicolumn{11}{c}{\bf \scriptsize impact energies $[$kev/amu$]$} \\
 & & \scriptsize 1.0 & \scriptsize 5.0 & \scriptsize 7.0 & \scriptsize 10.0 & \scriptsize 25.0 & \scriptsize 45.0 & \scriptsize 90. & \scriptsize 100. & \scriptsize 150. & \scriptsize 250. & \scriptsize 500. \\
\otoprule
 & & \multicolumn{11}{c}{\bf \scriptsize total cross sections [$\bf10^{-16}$ cm$\bf ^2$]}\\
 \multicolumn{2}{Sc}{CX} & 5.75E+02 & 4.86E+02 & 4.31E+02 & 3.41E+02 & 8.87E+01 & 2.08E+01 & 2.27E+00 & 1.52E+00 & 3.16E-01 & 3.01E-02 & 1.87E-03\\
 \multicolumn{2}{Sc}{ION} & 4.77E-01 &1.38E+00 & 1.85E+00 & 3.50E+00 & 2.13E+01 & 3.89E+01 & 5.37E+01 & 5.45E+01 & 4.11E+01 & 4.41E+01 & 2.30E+01\\
\otoprule
 & & \multicolumn{11}{Sc}{\bf \scriptsize state resolved CX cross sections [$\bf10^{-16}$ cm$\bf ^2$]}\\
\scriptsize $n$ & \scriptsize $\ell$ & \multicolumn{11}{c}{} \\
\midrule
   2 & & 1.00E-04& 6.54E-06& 1.99E-05& 4.54E-05& 7.66E-05& 4.15E-04& 2.26E-03& 2.14E-03& 1.91E-03& 6.00E-04& 1.87E-04 \\
\bottomrule
 \end{tabular}
 \label{t:c6hn2}
 \end{table}

\newpage
\section{References}

\bibliographystyle{jphysicsB}
\bibliography{nitro}

\end{document}